\documentclass[11pt,a4paper]{article}
\usepackage[utf8]{inputenc}
\usepackage{amsmath}
\usepackage{amsfonts}
\usepackage{amssymb}
\usepackage{amsthm,epsfig,epstopdf,titling,url,array,float}
\usepackage{multirow}
\usepackage{multicol}
\usepackage{geometry}
\usepackage{slashbox}
\usepackage{setspace}
\theoremstyle{plain}
\newtheorem{thm}{Theorem}[section]
\newtheorem{lem}[thm]{Lemma}

\usepackage{authblk}

\theoremstyle{definition}

\theoremstyle{remark}

\usepackage{mathtools}
\newcommand\givenbase[1][]{\:#1\lvert\:}
\let\given\givenbase

\DeclarePairedDelimiterX\Basics[1](){\let\given\sgiven #1}
\begin{document}

\title{An Improved Bound for Security in an Identity Disclosure Problem}
\author{Debolina Ghatak \hspace{7mm} and \hspace{7mm} Bimal K Roy}
\date{}
\maketitle
\doublespacing
\begin{abstract}
Identity disclosure of an individual from a released data is a matter of concern especially if it belongs to a category with low frequency in the data-set. Nayak et al. (2016) discussed this problem vividly in a census report and suggested a method of obfuscation, which would ensure that the probability of correctly identifying a unit from released data, would not exceed $\xi$ for some $\frac{1}{3}<\xi<1$. However, we observe that for the above method the level of security could be extended under certain conditions. In this paper, we discuss some conditions under which one can achieve a security for any $0<\xi<1$.
\end{abstract}
\section{Introduction}
\paragraph*{}
Many agencies release data to motivate statistical research and industrial work. But often these data-sets carry some information which may be sensitive to the individual bearing it. Erasing the name or some identity number associated with an individual may not always be sufficient to hide the identity of the individual. For example, imagine a situation where a data-set of $p$ variables corresponding to $n$ individuals are released and among these $p$ variables there is a variable named ``\emph{pin-code}"( sometimes called zip-code). Now ``\emph{pin-code}" is not supposed to be a sensitive variable, but it may happen that the intruder, who is trying to identify some individual in the data-set, has an idea about where the individual lives and thus can guess his ``\emph{pin-code}". In this case, if in the data-set there is no other individual having the same ``\emph{pin-code}", he can directly guess from this information which row in the data-set corresponds to the individual and thus the identity is revealed. Hence, suppressing identity numbers or names is not always sufficient to prevent identity disclosure. In case, there are a few variables with low frequency cells, it is usually easy for the intruder to identify the individual. 
\smallskip

Various articles including~\cite{BKP} \cite{TSBM} \cite{NCY} have discussed this problem and various authors have proposed different risk measures to evaluate the security in the released data. However,  here we follow the framework of Nayak et. al. \cite{NCY} where the intruder has a knowledge of the variable category $X_{(B)}$ corresponding to his target unit $B$. If the variable $X$ has $k$ categories $c_1, c_2, \ldots, c_k$, then we assume without loss of generality $X_{(B)} = c_1$ and the frequencies of the categories in the data-set are $T_1, T_2, \ldots, T_k$ respectively. 
\smallskip

If $T_1 = 1$, i.e. only $X_{(B)}$ has category $c_1$, the intruder can guess the row of his target unit with certainty. If $T_1$ is small, the intruder knows that his target unit is definitely one of the $T_1$ many units and then taking into consideration other information, he may successfully identify the row of his target unit or make a correct guess. Thus, in this case, the variable information must be suppressed before releasing the data. 

One way to do that is to completely erase the variable but that is not desirable to the statistician. The usual practice is to perturb the data in such a way so that the new data can be treated like the original data in making statistical inferences.    

If $\{X_1,X_2,\cdots,X_n\}$ is the original data-set and $\{Z_1,Z_2,\cdots,Z_n\}$ is the perturbed data then the transition matrix $P$ is given by, $((p_{ij}))$ where,
\begin{equation}
p_{ij}=P[Z=c_j|X=c_i] \mbox{ , $i,j=1,2,\cdots k$.}
\label{Eqn:P}
\end{equation} 
This matrix is not released and is unknown to the statistician. This method of obfuscation is known as the post-randomization method (PRAM). If we assume $\mathbf{T}=(T_1,T_2,\cdots, T_k) \sim Multinomial(\Pi_1,\Pi_2,\cdots,\Pi_k)$ then after transformation of $X$ to $Z$, if $\mathbf{S}=(S_1,S_2,\cdots, S_k)$ are the frequencies of each class $\{ c_1,c_2,\cdots, c_k \}$ in the perturbed data, then $\mathbf{S} \sim Multinomial({\Lambda}_1,{\Lambda}_2,\cdots,{\Lambda}_k)$, where $\mathbf{\Lambda}=P\mathbf{\Pi}$ ($\mathbf{\Lambda}:=({\Lambda}_1,{\Lambda}_2,\cdots,{\Lambda}_k)$, $\mathbf{\Pi}:=(\Pi_1,\Pi_2,\cdots,\Pi_k)$). If we want to treat $Z$ as the original data, we must have $\mathbf{\Pi}=\mathbf{\Lambda}=P\mathbf{\Pi}$. But $\mathbf{\Pi}$ is generally unknown to the one, who is masking the data. However, he can estimate $\mathbf{\Pi}$ from the original data with $\mathbf{T}/n$ where $n$ is the total sample size. If we want $\mathbf{S}/n$ to be an unbiased estimator of $\mathbf{\Pi}$, we must have,
\begin{equation}
E[\mathbf{S} \given \mathbf{T}]=\mathbf{T}/n \mbox{ , or equivalently, $P\mathbf{T}=\mathbf{T}$.}
\label{Eqn:PRAM}
\end{equation}
Gouweleeuw et.al. ( 1998) \cite{GKW} defined a post randomization method to be an invariant PRAM if $P$ satisfies Equation~\eqref{Eqn:PRAM}. The error due to estimation after post randomization was studied in the literature by various authors including Nayak et. al. \cite{NAS}.
\smallskip

One of the common techniques to achieve an invariant PRAM is to use an Inverse Frequency Post Randomization (IFPR) block diagonal matrix, in which the entire data-set is partitioned into few groups and within each group, categories are interchanged. If it is not desirable to change the category of some variable, it can be made to form its own block. Thus, if there are $m$ groups, given by $\{c_1,c_2,\cdots,c_{k_1}\}$, $\{c_{k_1+1},c_{k_1+2},\cdots,c_{k_1+k_2}\}$, $\ldots$, $\{c_{k_{m-1}+1},c_{k_{m-1}+2},\cdots,c_{k_{m-1}+k_m}\}$, where $k_1+k_2+\cdots +k_m=k$, then $p_{ij}>0$ if $c_j$ and $c_i$ fall into the same group and $p_{ij}=0$ if $c_j$ and $c_i$ fall into different groups. Within each group, $p_{ij}$ is given by, 
\begin{equation}
p_{ij} = 
\begin{cases}
1 - \theta / T_i & \mbox{ if } i = j \\
\frac{\theta}{(k^{\prime}-1)T_i} &  \mbox{ if } i \neq j 
\end{cases} \mbox{ , }
\label{Eqn:IFPR}
\end{equation}

where $0<\theta<1$ and $k^{\prime}>1$ is the block size of the group that $i$ and $j$ fall into.
\smallskip
However, the parameter $\theta$ of the model should be carefully chosen to ensure that the perturbed data is secured from the intruder, at least, up to a certain extent. To measure the risk of disclosure, Nayak et.al.~\cite{NCY} suggested checking whether the probability of correctly identifying an individual given any structure of $\mathbf{T}$ and any value of $S_1$ is bounded by some specified quantity $0<\xi<1$. Moreover, they showed that there exists a $\theta^{\star}$, where $0<\theta^{\star}<1$ which gives the transition matrix, $P(\theta^{\star})=((p^{\star}_{ij}))_{ 1\leq i \leq k, 1\leq j \leq k }$ where $p^{\star}_{ij}$ is chosen according to Equation~\eqref{Eqn:IFPR} with $\theta=\theta^{\star}$ for each $i,j=1,2,\cdots,k_1$ and $k_1$ is the  block size of the group $c_1$ belongs to. Without loss of generality, we assume the block $c_1$ belongs to is the first block. This matrix $P(\theta^{\star})$ when used to post randomize $X$,
\begin{equation}
P[ \mbox{ CM } \given S_1=a,T=\mathbf{t}] \leq \xi \mbox{ $\forall$ $a \geq 0$, $\forall t$,}
\label{Eqn:bound}
\end{equation}
for any $\frac{1}{3}  \leq \xi<1$, where CM denotes ``Correct Match". However, if we can extend the search range of $\theta$ from $0<\theta<1$ to $0<\theta<T_1$ and can find all categories in the first block that satisfy $T_j \geq T_1$ for all $j \neq 1$, then the level of security can be extended to any $0<\xi<1$. Note that, under this definition, there is no harm in the range of the probabilities as they certainly lie between 0 and 1. However, smaller the value of $\xi$, larger the block size is required. Therefore we can extend the security as far as the frequency distribution permits.
\section{Our Approach}
As mentioned earlier, our framework is similar to that of Nayak et.al. \cite{NCY}. From the intruder's point of view, we assume that as he gets access of the released data $\{Z_1,Z_2,\cdots, Z_n\}$, he checks the rows for which $Z_i=c_1$ for $\{i=1,2,\cdots,n \}$. Let $S_1$ be the total number of units having class $c_1$. If $S_1=0$, intruder stops searching for his target unit $B$ in the data-set. If $S_1=a$ for some $a>0$, he selects one unit randomly among these $a$ individuals and concludes that to be his target unit $B$. Under this assumption, we discuss how to choose the parameter $\theta$ of the IFPR block diagonal matrix ( See Equation~\eqref{Eqn:IFPR}), depending on $T_1$, so that the probability of correctly identifying unit $B$ is less than some specified $0<\xi<1$. Our method is described in the following paragraph.

Fix a $0<\xi<1$. Note that, if $T_1>\frac{1}{\xi}$, then there is no need for obfuscation as the intruder can choose one unit randomly and conclude it as his target unit $B$. Since, in the original data, the probability of correctly identifying $B$ is $1/T_1$, if $T_1>\frac{1}{\xi}$, the probability is less than $\xi$. This is quite intuitive since identification risk is a problem associated with low-frequency classes. If $T_1\leq \frac{1}{\xi}$, then we find $k_1=\mathcal{K}_1(\xi, T_1)$ classes ( where the function $\mathcal{K}_1$ is discussed in Sec. 3 ) such that for each of these classes $\{c_1,c_2,\cdots,c_{k_1}\}$, $T_j\geq T_1$ for each $j \in \{1,2,\cdots,k_1\}$. Such an event is usually feasible for moderate values of $\xi$ as $T_1$ usually has small values. If such classes are available, we can have any desired level of security, i.e., for any fixed $0<\xi<1$, there exists a corresponding $\theta^{\star}$ such that if the data is perturbed with matrix $P(\theta^{\star})$, Equation~\eqref{Eqn:bound} holds. If, however, such classes are not available, we can find the integer $n^{\star}$ such that $\frac{1}{n^{\star}}\leq \xi < \frac{1}{n^{\star}-1}$. Since $k_1$ classes are not available such that $T_j \geq T_1$ for each $j \in \{1,2,\cdots,k_1\}$, we now set $\xi_1=\frac{1}{n^{\star}-1}$ and try to find $k_1^1=\mathcal{K}_1(T_1,\xi_1)$ classes such that $T_j \geq T_1$ for each $j \in \{1,2,\cdots,k_1^1\}$. If we fail, we next try for $\xi_2=\frac{1}{n^{\star}-2}$ and so on until we get a success for some $\xi_l=\frac{1}{n^{\star}-l}$. 
 Since for $\xi=\frac{1}{n^{\star}-l}$, there exists $k_1^l=\mathcal{K}_1(T_1,\xi_l)$ classes such that $T_j \geq T_1$ for each $j \in \{1,2,\cdots,k_1^l\}$, and a $\theta^{\star}$, such that if the data is perturbed with $P(\theta^{\star})$, then Equation~\eqref{Eqn:bound} is satisfied for any $\frac{1}{n^{\star}-l}<\xi<1$. According to Nayak et. al. \cite{NCY}, there is always a solution for $\xi \geq \frac{1}{3}$ which implies $n^{\star}$ can take a minimum value. However, $n^{\star}$ can take higher values in many cases.
 
 \section{Model,Assumptions and Results}
 
 As discussed earlier, the goal of the paper is to find out a method by which a data can be perturbed ensuring as much security as possible. Since security is an abstract term, we limit ourselves to ensure that the measure, given by Equation~\eqref{Eqn:bound}) holds for low values of $\xi$. Smaller the value of $\xi$, better the security of the data.
 \smallskip
 Let us denote, by $R_1(a,\mathbf{t})$, the probability of correctly identifying the individual from released data given $S_1=a$ and the frequency distribution of $X$ given by $\mathbf{t}:=(t_1,t_2,\cdots,t_k)$. In other words,
 \begin{equation} \label{eqn:R_1}
 R_1(a, t) = P[CM \given S_1 = a, T = \mathbf{t}], ~~ a \geq 0, t \in \mathbb{R}^k.
 \end{equation}
 If $R_1(a, \mathbf{t})$ is bounded by $\xi$ for any $\mathbf{t}$, then note that 
 \begin{equation}
 R_1(a) = P[CM \given S_1 = a],
 \label{Eqn:R_1(2)}
 \end{equation}
 is bounded by $\xi$ for any $ a \geq 0$, which signifies that the probability of correctly identifying an individual is less than $\xi$, no matter how small or large the frequency of category $c_1$ is, in the released data. $R_1(a, \mathbf{t})$ is used instead of $R_1(a)$ because it is hard to calculate the probability if $\mathbf{t}$ is not known.  Note that, CM stands for ``Correct Match" in the above equations~\eqref{eqn:R_1}~\eqref{Eqn:R_1(2)}.
 \smallskip
 
 Recall that if we use, IFPR block diagonal matrix to perturb $X$, the category $c_1$ may get changed to one of $\{c_1, c_2, \ldots, c_{k_1}\}$, $k_1 \geq 2$ with positive probability. let us denote $\alpha_i = p_{1i}, \beta_i = \frac{\alpha_i}{1 - \alpha_i}$ for $i \in \{1, 2, \ldots, k_1\}$. Observe that, $R_1(a, \mathbf{t})$ can be re-written as
\begin{eqnarray*}
R_1(a,\mathbf{t}) = & P[CM \given S_1=a,Z_{(B)}=c_1,T=\mathbf{t}]P[Z_{(B)}=c_1 \given S_1=a, T=\mathbf{t}] \\
&  +P[CM \given S_1=a,Z_{(B)} \neq c_1,T=\mathbf{t}]P[Z_{(B)} \neq c_1 \given S_1=a, T=\mathbf{t}].
\end{eqnarray*}

 By our assumption, since the intruder searches his target unit $B$ among the ones with category $c_1$, $P[CM \given S_1=a,Z_{(B)} \neq c_1,T=\mathbf{t}]=0$. Again, since, the intruder is assumed to choose randomly one unit among $a$ units to be $B$, $P[CM \given S_1=a,Z_{(B)}= c_1,T=\mathbf{t}]=\frac{1}{a}$ for any $\mathbf{t}$. Thus, 
 
 \begin{equation}
 R_1(a,\mathbf{t})=\frac{1}{a}P[Z_{(B)}=c_1 \given S_1=a, T=\mathbf{t}].
 \label{eqn:R_1_main}
 \end{equation}
 
 Again, we have, 

\begin{eqnarray}
P[Z_{(B)}=c_1,S_1=a,\given T=\mathbf{t}] = \alpha_1 \sum \prod_{i=1}^{k_1}{{T_i^{\star}}\choose{a_i}}\alpha_i^{a_i}(1-\alpha_i)^{T_i^{\star}-a_i} \nonumber \\
\hspace*{18mm}= \alpha_1 [\prod_{i=1}^{k_1}{(1-\alpha_i)^{T_i^{\star}}}] \sum \prod_{i=1}^{k_1}{{{T_i^{\star}}\choose{a_i}}\beta_i^{a_i}}
\label{eqn:term_1}
\end{eqnarray}

where $T_1^{\star}=T_1-1$, $T_i^{\star}=T_i$, $i \geq 2$ and the sum is over all integer-valued $a_1,a_2,\cdots a_{k_1}$ such that $0 \leq a_i \leq T_i^{\star}$ and $\sum{a_i}=a-1$. We denote the sum by $\Sigma_{a-1}$
\begin{equation}
P[Z_{(B)} \neq c_1,S_1=a,\given T=\mathbf{t}]=(1-\alpha_1)[\prod_{i=1}^{k_1}{(1-\alpha_i)^{T_i^{\star}}}]\Sigma_{a}
\label{eqn:term_2}
\end{equation}
Equation~\eqref{eqn:term_1} and \eqref{eqn:term_2} implies that  $$P[S_1=a \given T=\mathbf{t}]=\prod_{i=1}^{k_1}{(1-\alpha_i)^{T_i^{\star}}}(\alpha_1\Sigma_{a-1}+(1-\alpha_1)\Sigma_a)$$ and since $$ P[Z_{(B)} = c_1 \given S_1=a, T=\mathbf{t}]=\frac{P[Z_{(B)}=c_1,S_1=a \given T=\mathbf{t}]}{P[S_1=a \given T=\mathbf{t}]} $$
from Equation~\eqref{eqn:R_1_main} , we finally have,
\begin{eqnarray}
R_1(a,\mathbf{t}) &=\frac{1}{a}\left[\frac{\alpha_1\Sigma_{a-1}}{\alpha_1\Sigma_{a-1}+(1-\alpha_1)\Sigma_a}\right]  \nonumber \\  &=\frac{1}{a}\left[1+\frac{1}{\beta_1}\frac{\Sigma_a}{\Sigma_{a-1}}\right]^{-1}.
\label{eqn:R_1_final}
\end{eqnarray}

Nayak et.al. \cite{NCY} observed that although it seems intuitive that $R_1(1,t) \geq R_1(a,\mathbf{t})$ for any $t$, $a>1$ there are certain cases it does not hold true. However, they proved that if $\alpha_1 \geq \alpha_j$, i.e., $\beta_1 \geq \beta_j$ for all $j=1,2,\cdots,k_1$, then $R_1(1,t) \geq R_1(2,t)$ for any $t$. Intuitively, if $\beta_1$ is highest, i.e., the odds that $c_1$ goes to any category other than  $c_1$, then the risk of disclosure should be maximum if $a=1$. We checked that this is quite true which leads us to our first result, stated in the following theorem and the proof is given in Appendix Section.

\begin{thm}
If $\alpha_1 \geq \alpha_j$, i.e., $\beta_1 \geq \beta_j$ for any $j=1,2,\cdots,k_1$, then $R_1(1,t) \geq R_1(a,\mathbf{t})$ for any $t$, $a>1$, where $R_1(a,\mathbf{t})$ is given by Equation \eqref{eqn:R_1_final}.
\label{Result:One}
\end{thm}
Assuming Theorem \ref{Result:One} holds, proving Equation~\eqref{Eqn:bound} is equivalent to prove that $R_1(1,t) \leq \xi$ for any $t$. For this condition to hold, we must carefully choose the parameter $\theta$ in \eqref{Eqn:P}. Due to Nayak et. al. \cite{NCY}, we have,

\begin{eqnarray}
R_1(1,T) = \left[ T_1+ \frac{\theta}{T_1- \theta}\sum_{i=2}^{k_1}{\frac{\theta T_i}{(k_1-1)T_i-\theta}} \right]^{-1} \nonumber \\
=(T_1-\theta)\left[ T_1(T_1-\theta)+ \theta^2 \sum_{i=2}^{k_1}{\frac{T_i}{(k_1-1)T_i-\theta}} \right]^{-1} \nonumber \\
 \leq \frac{T_1-\theta}{T_1(T_1-\theta)+\theta^2} =\psi(T_1,\theta)
\end{eqnarray}

To proceed further we also need the following lemma, proof of which is deferred in Appendix Section.

\begin{lem}
For any fixed $0<\xi<1$, there exists a $\theta^{\star} \in (0,T_1)$ such that $\psi(\theta,T_1) \leq \xi$.
\label{Lem:One}
\end{lem}
For Theorem \ref{Result:One} to hold, in an IFPR block diagonal matrix, we must have $\frac{T_1-\theta}{\theta} \geq \frac{\theta}{(k_1-\theta)T_1-\theta}$ which leads to the condition, $\theta \leq \frac{T_1}{1+\frac{T_1}{T_j(k_1-1)}}$, i.e., $k_1-1 \geq \frac{\theta}{T_1-\theta}\frac{T_1}{T_j}$. Note that, if $k_1-1 \geq \frac{\theta}{T_1-\theta}$, and $\frac{T_1}{T_j} \leq 1$, $k_1-1 \geq \frac{\theta}{T_1-\theta}\frac{T_1}{T_j}$. Hence, it is enough to find $\mathcal{K}(\theta,T_1)=1+\frac{\theta}{T_1-\theta}=\frac{T_1}{T_1-\theta}$ for Theorem \ref{Result:One} to hold. Again, $\theta$ is chosen by solving $\psi(\theta,T_1)=\xi$. Thus, for fixed $\xi$ and $T_1$ we have a $\theta$ and a corresponding $\mathcal{K}_1(\xi,T_1)$ which is the largest integer contained in $\mathcal{K}(\theta,T_1)$. $\mathcal{K}_1(\xi,T_1)$ is the minimum number of categories required to form the block containing $c_1$. For some possible choices of $\xi$ and some possible values of $T_1$, the value of $\mathcal{K}_1(\xi,T_1)$ is calculated and given in Table \ref{Table:K1}. While choosing the block size, one must note that the block size $k_1$ must be larger than or at least equal to $\mathcal{K}_1(\xi,T_1)$ to ensure Equation \eqref{Eqn:bound}.

\begin{table}
\caption{ Showing minimum block size required for some possible choices of  security level $\xi$ and some possible values of class frequency $T_1$}
\begin{center}
\begin{tabular}{|c||c|c|c|c|c|c|c|}
\hline
\backslashbox{$T_1$}{$\xi$} & 0.1 & 0.125 & 0.15 & 0.175 & 0.2 & 0.25 & 0.3 \\
\hline
\hline
1 & 11 & 9 & 8 & 7 & 6 & 5 & 5 \\
\hline
2 & 6 & 5 & 5 & 4 & 4 & 3 & 3 \\
\hline
3 & 5 & 4 & 3 & 3 & 3 & 2 & 2 \\
\hline
4 & 4 & 3 & 3 & 2 & 2 & 2 & 2 \\
\hline
5 & 3 & 3 & 2 & 2 & 2 & 2 & 2 \\
\hline
6 & 3 & 2 & 2 & 2 & 2 & 2 & 2 \\
\hline
7 & 2 & 2 & 2 & 2 & 2 & 2 & 2 \\
\hline
8 & 2 & 2 & 2 & 2 & 2 & 2 & 2 \\
\hline
9 & 2 & 2 & 2 & 2 & 2 & 2 & 2 \\
\hline
10 & 2 & 2 & 2 & 2 & 2 & 2 & 2 \\
\hline
\end{tabular}
\end{center}
\label{Table:K1}
\end{table}

\section{Simulation Results}
To illustrate the process, we simulate a sample of size $n=2000$ from $k=8$ categories such that the probability of falling into a category is given by the vector $\mathbf{\Pi}=(0.001,0.1,0.2,0.05,0.12,0.13,0.301,0.098)$. The sample has frequency distribution given by Table \ref{Table:T}.

\begin{table}[H]
\centering
\caption{Table showing frequencies of Categories for True Data from Simulated data-set}
\begin{tabular}{|c|c|}
\hline
Category & T \\
\hline
1 &  2 \\
2 & 205 \\
3 & 431 \\
4 & 106\\
5 & 230 \\
6 & 221 \\
7 & 611 \\
8 & 194 \\
\hline
\end{tabular}
\label{Table:T}
\end{table}

Two units in the data-set have Category 1, one of which is unit $B=780$. Since $T_1=2$, the probability of Correct Match from true data is 0.5 which is very high. We want this probability to be lower, say below $\xi=0.1$. So, we transform the data to $Z$ using the IPRAM method with a transition matrix $P$. To choose an ideal $P$ we apply the procedure of this paper. From Table \ref{Table:K1}, we get the required block size is 6. So, we would apply transition to the first $k_1=6$ categories with the lowest probability of occurrence and do not alter the categories for the rest 2 categories. To solve for $h(\theta)=\xi$, we have $\theta^{\star}=1.656854$ which gives the transition matrix,

$$P=\left(\begin{matrix} 
 0.172 & 0.166 & 0 & 0.166 & 0.166 & 0.166 & 0 & 0.166 \\
 0.002 & 0.992 & 0 & 0.002 & 0.002 & 0.002 & 0 & 0.002 \\
  0 & 0 & 1 & 0 & 0 & 0 & 0 & 0 \\
  0.003 & 0.003 & 0 & 0.984 & 0.003 & 0.003 & 0 & 0.003 \\
  0.001 & 0.001 & 0 & 0.001 & 0.993 & 0.001 & 0 & 0.001 \\
  0.001 & 0.001 & 0 & 0.001 & 0.001 & 0.993 & 0 & 0.001 \\
  0 & 0 & 0 & 0 & 0 & 0 & 1 & 0 \\
  0.002 & 0.002 & 0 & 0.002 & 0.002 & 0.002 & 0 & 0.991 
\end{matrix}\right)$$
Using this transition matrix we ran 1000 simulations to get 1000 different $Z$s. The mean squared estimation error for each category is given by $E=(4.9350e-07, 7.6125e-07, 0.0000e+00, 7.4300e-07, 8.8550e-07, 7.8375e-07, 0.0000e+00, 8.5550e-07)$ which is quite low and the average probability of correct match in 1000 simulations is $0.07639286<0.1$.

The process thus seems to work well for simulated data.
\section{Conclusion}

The method works fine in most practical cases, because, in general, since we want to obfuscate categories with low frequency, there will be sufficient number of categories with higher frequency values than them. Accordingly, the security level can be increased. 

However, the greatest drawback of this method of obfuscation is that we have assumed the game of the intruder, i.e., it selects one of the units with the desired categorical value randomly looking at the obfuscated data. But this is not expected to happen since in most cases there will be many regressive variables associated and the selection will not be, in general, random. This problem was also discussed in \cite{TSBM}. 

However, if the model assumptions hold true, the discussed method is successful in giving a better security. 

\section*{Appendix}
\subsection*{\underline{Proof to Theorem \ref{Result:One}}}

To prove the result, we need to show $R_1(a+1,T) \leq R_1(1,T)$, i.e., $\frac{1}{a+1}(1+\frac{\Sigma_{a+1}}{\Sigma_a\beta_1})^{-1} \leq (1+\frac{\Sigma}{\beta_1})^{-1}$ which leads us to check an equivalent statement, 
\begin{equation}
\tilde{\Sigma}_{a+1} - \Sigma\tilde{\Sigma}_a + \beta_1 \tilde{\Sigma}_a \geq 0
\label{ineq:criterion}
\end{equation}
where $\tilde{\Sigma}_a=a!\Sigma_a$ ($\Sigma_a$ as defined in Equation (\ref{eqn:term_1}) and (\ref{eqn:term_2})). Thus, we will need to check if \ref{ineq:criterion} holds for all $a$ and all $k_1$ to prove Theorem \ref{Result:One}.
\smallskip

\noindent
We will prove this result by a two dimensional induction procedure. First, we show that the statement is true for $k_1=2$ for all $a \in \mathbb{N}$, then we show that if the statement is true for $k_1=k_{1_0}$, then it is true for $k_1=k_{1_0}+1$ for all $a$.
\smallskip

\noindent
\textbf{\underline{Case: $k_1=2$}:} Since, $\Sigma_1=\sum{T^\star_{i}\beta_i}$ and

$$ \tilde{\Sigma}_a=\sum_{s=0}^a{ {a\choose s} \sum_{i_1 \neq i_2}{T^\star_{i_1}(T^\star_{i_1}-1)\cdots(T^\star_{i_1}-s+1)T^\star_{i_2}(T^\star_{i_2}-1)\cdots(T^\star_{i_2}-\overline{a-s})\beta_{i_1}^s\beta_{i_2}^{a-s}}} $$

We have, 

\paragraph*{}
$\begin{array}{ll}
 \tilde{\Sigma}_a\Sigma_1 & =\sum_{s=0}^a{ {a\choose s} \sum_{i_1 \neq i_2}{{T^\star_{i_1}}^2(T^\star_{i_1}-1)\cdots(T^\star_{i_1}-s+1)T^\star_{i_2}(T^\star_{i_2}-1)\cdots(T^\star_{i_2}-\overline{a-s})\beta_{i_1}^{s+1}\beta_{i_2}^{a-s}}}\\
  &   + \sum_{s=0}^a{ {a\choose s} \sum_{i_1 \neq i_2}{T^\star_{i_1}(T^\star_{i_1}-1)\cdots(T^\star_{i_1}-s+1){T^\star_{i_2}}^2(T^\star_{i_2}-1)\cdots(T^\star_{i_2}-\overline{a-s})\beta_{i_1}^s\beta_{i_2}^{a-s+1}}}
\end{array}$
\paragraph*{}
Writing $\Sigma_{a+1}$ similarly, we note that there are $a+2$ terms in the expansion of $\tilde{\Sigma}_{a+1} - \Sigma\tilde{\Sigma}_a + \beta_1 \tilde{\Sigma}_a$.
\paragraph*{}

$\begin{array}{ll}
\mbox{First term} &= {{a+1}\choose 0}\sum_{i_2=1}^{k_1}{T^\star_{i_2}(T^\star_{i_2}-1)\cdots(T^\star_{i_2}-a)\beta_{i_2}^{a+1}}-{a \choose 0}\sum_{i_2=1}^{k_1}{{T^\star_{i_2}}^2(T^\star_{i_2}-1)\cdots(T^\star_{i_2}-a+1)\beta_{i_2}^{a+1}} \\
&  \hspace{20mm} + a{a\choose 0} \sum_{i_2=1}^{k_1}{T^\star_{i_2}(T^\star_{i_2}-1)\cdots(T^\star_{i_2}-a+1)\beta_{i_2}^{a}\beta_1} \\
&=a\sum_{i_2=1}^{k_1}{T^\star_{i_2}(T^\star_{i_2}-1)\cdots(T^\star_{i_2}-a+1)\beta_{i_2}^{a}(\beta_1-\beta_{i_2})}
\end{array}$
\paragraph*{}
For $s=1,2,\cdots a$,
\paragraph*{}
\begin{footnotesize}
$\begin{array}{ll}
\mbox{$(s+1)^{th}$ term} &= {{a+1}\choose s}\sum_{i_1 \neq i_2}{T^\star_{i_1}(T^\star_{i_1}-1)\cdots(T^\star_{i_1}-s+1){T^\star_{i_2}}(T^\star_{i_2}-1)\cdots(T^\star_{i_2}-\overline{a+1-s}+1)\beta_{i_1}^s\beta_{i_2}^{a+1-s}} \\
& - { a\choose s}\sum_{i_1 \neq i_2}{T^\star_{i_1}(T^\star_{i_1}-1)\cdots(T^\star_{i_1}-s+1){T^\star_{i_2}}^2(T^\star_{i_2}-1)\cdots(T^\star_{i_2}-\overline{a-s}+1)\beta_{i_1}^s\beta_{i_2}^{a+1-s}} \\
& - {{a}\choose {s-1}}\sum_{i_1 \neq i_2}{{T^\star_{i_1}}^2(T^\star_{i_1}-1)\cdots(T^\star_{i_1}-s+2){T^\star_{i_2}}(T^\star_{i_2}-1)\cdots(T^\star_{i_2}-\overline{a-s})\beta_{i_1}^s\beta_{i_2}^{a+1-s}}\\
& + a\beta_1{{a}\choose s}\sum_{i_1 \neq i_2}{T^\star_{i_1}(T^\star_{i_1}-1)\cdots(T^\star_{i_1}-s+1){T^\star_{i_2}}(T^\star_{i_2}-1)\cdots(T^\star_{i_2}-\overline{a-s}+1)\beta_{i_1}^s\beta_{i_2}^{a-s}} \\
& \\
&={{a}\choose s}\sum_{i_1 \neq i_2}{T^\star_{i_1}(T^\star_{i_1}-1)\cdots(T^\star_{i_1}-s+1){T^\star_{i_2}}(T^\star_{i_2}-1)\cdots(T^\star_{i_2}-\overline{a-s}+1)(-\overline{a-s})\beta_{i_1}^s\beta_{i_2}^{a+1-s}} \\
&-{{a}\choose {s-1}}(s-1)\sum_{i_1 \neq i_2}{T^\star_{i_1}(T^\star_{i_1}-1)\cdots(T^\star_{i_1}-s+2){T^\star_{i_2}}(T^\star_{i_2}-1)\cdots(T^\star_{i_2}-\overline{a-s})\beta_{i_1}^s\beta_{i_2}^{a+1-s}} \\
& + a\beta_1{{a}\choose s}\sum_{i_1 \neq i_2}{T^\star_{i_1}(T^\star_{i_1}-1)\cdots(T^\star_{i_1}-s+1){T^\star_{i_2}}(T^\star_{i_2}-1)\cdots(T^\star_{i_2}-\overline{a-s}+1)\beta_{i_1}^s\beta_{i_2}^{a-s}} \\
& \\
& [\mbox{Using Pascal's rule ${{a+1}\choose s}$=${a \choose s} +{a \choose {s-1}}$}] \\
& \\
&=a{{a}\choose s}\sum_{i_1 \neq i_2}{T^\star_{i_1}(T^\star_{i_1}-1)\cdots(T^\star_{i_1}-s+1){T^\star_{i_2}}(T^\star_{i_2}-1)\cdots(T^\star_{i_2}-\overline{a-s}+1)\beta_{i_1}^s\beta_{i_2}^{a-s}}(\beta_1-\beta_{i_2}) \\
& + {{a}\choose s}s\beta_1\sum_{i_1 \neq i_2}{T^\star_{i_1}(T^\star_{i_1}-1)\cdots(T^\star_{i_1}-s+1){T^\star_{i_2}}(T^\star_{i_2}-1)\cdots(T^\star_{i_2}-\overline{a-s}+1)\beta_{i_1}^s\beta_{i_2}^{a-s}} \\
&- {{a}\choose {s-1}}(s-1)\sum_{i_1 \neq i_2}{T^\star_{i_1}(T^\star_{i_1}-1)\cdots(T^\star_{i_1}-s+2){T^\star_{i_2}}^2(T^\star_{i_2}-1)\cdots(T^\star_{i_2}-\overline{a-s})\beta_{i_1}^s\beta_{i_2}^{a+1-s}} \\
& \\
& \geq  {{a}\choose s}s\beta_1\sum_{i_1 \neq i_2}{T^\star_{i_1}(T^\star_{i_1}-1)\cdots(T^\star_{i_1}-s+1){T^\star_{i_2}}(T^\star_{i_2}-1)\cdots(T^\star_{i_2}-\overline{a-s}+1)\beta_{i_1}^s\beta_{i_2}^{a-s}} \\
&- {{a}\choose {s-1}}(s-1)\sum_{i_1 \neq i_2}{T^\star_{i_1}(T^\star_{i_1}-1)\cdots(T^\star_{i_1}-\overline{s-1}+1){T^\star_{i_2}}^2(T^\star_{i_2}-1)\cdots(T^\star_{i_2}-\overline{a-(s-1)}+1)\beta_{i_1}^s\beta_{i_2}^{a-(s-1)}} \\
& \mbox{ [Since, $\beta_1 \geq \beta_i \forall i$]}
\end{array}$
\end{footnotesize}
\paragraph*{}
In the last expression, let us denote the first term by $Term(s,\beta1)$ and the second term by $Term(s-1,\beta)$. Note that since $\beta_1 \geq \beta_i \forall i$ $Term(s,\beta_1)-Term(s,\beta) \geq 0$.

$$
\mbox{$(a+2)^{th}$ term} = {{a+1}\choose{a+1}}\sum_{i_1=1}^{k_1}{T^\star_{i_1}(T^\star_{i_1}-1)\cdots(T^\star_{i_1}-a)\beta_{i_1}^{a+1}}-{a \choose a}\sum_{i_1=1}^{k_1}{{T^\star_{i_1}}^2(T^\star_{i_1}-1)\cdots(T^\star_{i_1}-a+1)\beta_{i_1}^{a+1}} 
$$

Thus, it can be clearly seen that,
\paragraph*{}
$
\begin{array}{ll}
&\tilde{\Sigma}_{a+1} - \Sigma_1\tilde{\Sigma}_a + \beta_1 \tilde{\Sigma}_a \\
&\geq Term(1,\beta_1)+Term(2,\beta_1)-Term(1,\beta)+Term(3,\beta_1)-Term(2,\beta) \\
&+\cdots +Term(a,\beta_1)-Term(a-1,\beta) +({{a+1}\choose{a+1}}\sum_{i_1=1}^{k_1}{T^\star_{i_1}(T^\star_{i_1}-1)\cdots(T^\star_{i_1}-a)\beta_{i_1}^{a+1}})-Term(a,\beta) \\
&\geq 0 
\end{array}
$
\paragraph*{}
Hence, (\ref{ineq:criterion}) is true for $k_1=2$ for any $a$. Now, let it be true for some $k_1=k_{1_0}$ ,  $k_{1_0} \in \{2,3, \dots\}$. We will show then that (\ref{ineq:criterion}) is true for $k_1=k_{1_0}+1$.
\vspace{12mm}

\noindent
\textbf{\underline{Case: $k_1=k_{1_0}+1$}:} The general expression for $\tilde{\Sigma}_a$ can be given by the following expression.

$$ \tilde{\Sigma}_a =\sum_{a_1+a_2+\cdots + a_{k_1}=a} {\frac{a!}{a_1!\cdots a_{k_1}!}}\sum_{i_1 \neq i_2 \cdots \neq i_{k_1}}{T^{\star}_{i_1}\ldots (T^{\star}_{i_1}-a_1+1) \ldots T^{\star}_{i_{k_1}}\ldots (T^{\star}_{i_{k_1}}-a_{k_1}+1) \beta_{i_1}^{a_1}\beta_{i_2}^{a_2}\ldots \beta_{i_{k_1}}^{a_{k_1}}}$$
Since for any $\{a_1,a_2,\cdots,a_{k_1} \geq 0 , \sum_{i=1}^{k_1}{a_i}=a : \frac{a!}{a_1!a_2!\cdots a_{k_1}!}=\frac{a!}{a_!!(a-a_1)!}\frac{(a-a_1)!}{a_2!a_3!\cdots a_{k_1}!}\}$, we can write,
$$\tilde{\Sigma}_{a}= \sum_{s=0}^{a}{\frac{a!}{s!(a-s)!}\sum_{i_1=1}^{k_1}{T^\star_{i_1}(T^\star_{i_1}-1)\cdots(T^\star_{i_1}-s+1)\beta_{i_1}^{s}\tilde{\Sigma}_{(a-s,k_{1_0})}}}$$
where $\tilde{\Sigma}_{(s,k_{1_0})}=s!\Sigma_s$ for $k_{1_0}$ categories instead of $k_1=k_{1_0}+1$ categories. Like before, we write down the terms of $\tilde{\Sigma}_{a+1} - \Sigma\tilde{\Sigma}_a + \beta_1 \tilde{\Sigma}_a$.
\paragraph*{}
$\begin{array}{ll}
\mbox{First term} &=\tilde{\Sigma}_{(a+1,k_{1_0})}-\sum_{i_1=1}^{k_1}{T_{i_1}^{\star}\beta_{i_1}\tilde{\Sigma}_{(a,k_{1_0})}}-\tilde{\Sigma}_{(a,k_{1_0})}\tilde{\Sigma}_{(1,k_{1_0})}+a\beta_1\tilde{\Sigma}_{(a,k_{1_0})} \\
&=(\tilde{\Sigma}_{(a+1,k_{1_0})}-\tilde{\Sigma}_{(a,k_{1_0})}\tilde{\Sigma}_{(1,k_{1_0})}+a\beta_1\tilde{\Sigma}_{(a,k_{1_0})})-\sum_{i_1=1}^{k_1}{T_{i_1}^{\star}\beta_{i_1}\tilde{\Sigma}_{(a,k_{1_0})}} \\
&\geq -\sum_{i_1=1}^{k_1}{T_{i_1}^{\star}\beta_{i_1}\tilde{\Sigma}_{(a,k_{1_0})}} \mbox{ \hspace{10mm}[by Assumption over size $k_{1_0}$]}\\
\end{array}$
\paragraph*{}
For $s=1,2,\cdots a$,
\paragraph*{}
\begin{footnotesize}
$\begin{array}{ll}
\mbox{$(s+1)^{th}$ term} &= {\frac{(a+1)!}{s!(a-s+1)!}\sum_{i_1=1}^{k_1}{T^\star_{i_1}(T^\star_{i_1}-1)\cdots(T^\star_{i_1}-s+1)\beta_{i_1}^{s}\tilde{\Sigma}_{(a-s+1,k_{1_0})}}} \\
&-{\frac{a!}{s!(a-s)!}\sum_{i_1=1}^{k_1}{{T^\star_{i_1}}^2(T^\star_{i_1}-1)\cdots(T^\star_{i_1}-s+1)\beta_{i_1}^{s+1}\tilde{\Sigma}_{(a-s,k_{1_0})}}}\\
& -({\frac{a!}{s!(a-s)!}\sum_{i_1=1}^{k_1}{T^\star_{i_1}(T^\star_{i_1}-1)\cdots(T^\star_{i_1}-s+1)\beta_{i_1}^{s}\tilde{\Sigma}_{(a-s,k_{1_0})}}})(\tilde{\Sigma}_{(1,k_{1_0})})\\
& +a\beta_1\frac{a!}{s!(a-s)!}\sum_{i_1=1}^{k_1}{T^\star_{i_1}(T^\star_{i_1}-1)\cdots(T^\star_{i_1}-s+1)\beta_{i_1}^{s}\tilde{\Sigma}_{(a-s,k_{1_0})}} \\
& \\
& = {a \choose s} \sum_{i_1=1}^{k_1}{T^\star_{i_1}(T^\star_{i_1}-1)\cdots(T^\star_{i_1}-s+1)\beta_{i_1}^{s}(\tilde{\Sigma}_{(a+1-s,k_{1_0})}}-\tilde{\Sigma}_{(a-s,k_{1_0})}\tilde{\Sigma}_{(1,k_{1_0})}+(a-s)\beta_1\tilde{\Sigma}_{(a-s,k_{1_0})}) \\
& + {a \choose {s-1}} \sum_{i_1=1}^{k_1}{T^\star_{i_1}(T^\star_{i_1}-1)\cdots(T^\star_{i_1}-s+1)\beta_{i_1}^{s}(\tilde{\Sigma}_{(a-\overline{s-1},k_{1_0})}} \\
& -{a \choose s} \sum_{i_1=1}^{k_1}{{T^\star_{i_1}}^2(T^\star_{i_1}-1)\cdots(T^\star_{i_1}-s+1)\beta_{i_1}^{s+1}(\tilde{\Sigma}_{(a-s,k_{1_0})}} \\
& + s\beta_1{a \choose {s}} \sum_{i_1=1}^{k_1}{T^\star_{i_1}(T^\star_{i_1}-1)\cdots(T^\star_{i_1}-s+1)\beta_{i_1}^{s}(\tilde{\Sigma}_{(a-s,k_{1_0})}} \mbox{ [ Using Pascal's rule]} \\
& \\
& \geq + {a \choose {s-1}} \sum_{i_1=1}^{k_1}{T^\star_{i_1}(T^\star_{i_1}-1)\cdots(T^\star_{i_1}-s+1)\beta_{i_1}^{s}(\tilde{\Sigma}_{(a-\overline{s-1},k_{1_0})}} \\
& -{a \choose s} \sum_{i_1=1}^{k_1}{{T^\star_{i_1}}^2(T^\star_{i_1}-1)\cdots(T^\star_{i_1}-s+1)\beta_{i_1}^{s+1}(\tilde{\Sigma}_{(a-s,k_{1_0})}} \\
& + s\beta_1{a \choose {s}} \sum_{i_1=1}^{k_1}{T^\star_{i_1}(T^\star_{i_1}-1)\cdots(T^\star_{i_1}-s+1)\beta_{i_1}^{s}(\tilde{\Sigma}_{(a-s,k_{1_0})}} \mbox{ [ Using Assumption on size $k_{1_0}$]} 
\end{array}$

\end{footnotesize}

$$ \mbox{$(a+2)^{th}$ term}=\sum_{i_1=1}^{k_1}{T^\star_{i_1}(T^\star_{i_1}-1)\cdots(T^\star_{i_1}-a)\beta_{i_1}^{a+1}}$$
\paragraph*{}
Summing all the elements we get, 
\paragraph*{}
$\begin{array}{ll}
&\tilde{\Sigma}_{a+1} - \Sigma\tilde{\Sigma}_a + \beta_1 \tilde{\Sigma}_a \\
&={a \choose 1}\sum_{i_1=1}^{k_1}{T^\star_{i_1}\beta_{i_1}\tilde{\Sigma}_{(a-1,k_{1_0})}(\beta_1-\beta_{i_1})}+{a \choose 2}\sum_{i_1=1}^{k_1}{T^\star_{i_1}(T^\star_{i_1}-1)\beta_{i_1}^2\tilde{\Sigma}_{(a-2,k_{1_0})}(\beta_1-\beta_{i_1})} \\
& + \cdots + {a \choose a}\sum_{i_1=1}^{k_1}{T^\star_{i_1}(T^\star_{i_1}-1)\cdots(T^\star_{i_1}-a+1)\beta_{i_1}^a(\beta_1-\beta_{i_1})} \geq 0 \mbox{ [ Since $\beta_1 \geq \beta_i \forall i$]}
\end{array}$
\paragraph*{}
Thus the statement is true for $k_1=k_{1_0}+1$ if true for $k_{1_0}$ for any $a \geq 1$. Thus, we see (\ref{ineq:criterion}) always holds and hence the proof. 
\subsection*{\underline{Proof to Lemma \ref{Lem:One}}}
For $T_1 \geq 2$, $\psi(1,\theta) = \psi(T_1,\theta)$ iff $\frac{1-\theta}{1-\theta+\theta^2} = \frac{T_1-\theta}{T_1(T_1-\theta)+\theta^2}$, i.e., $\theta = \frac{T_1}{T_1+1}$. Consider, 
$$ h(\theta) = 
\begin{cases}
\psi(1,\theta) & ,\mbox{ if } \theta<\frac{T_1}{T_1+1} \\
 \psi(T_1,\theta) & ,\mbox{ if } \theta \geq \frac{T_1}{T_1+1}
\end{cases} 
$$
Note that, $h(\theta)$ is continuous and strictly decreasing in $\theta \in (0,1)$ with $h(0)=1$, $h(T_1)=0$. By Mean Value Theorem, there must exist a $\theta^{\star} \in (0,T_1)$ such that $h(\theta^{\star})=\xi$ for $0<\xi<1$.
\end{document}